\documentclass[aps,prd,nofootinbib,preprint,superscriptaddress]{revtex4}  
\usepackage{amsmath,amssymb,exscale} 
\usepackage{psfrag,graphicx}
\usepackage{epsfig}
\usepackage{color,amsmath,amscd,shadow,fancybox,axodraw}
\def\beq{\begin{equation}}
\def\eeq{\end{equation}}
\def\bea{\begin{eqnarray}}
\def\eea{\end{eqnarray}}
\renewcommand{\le}{\ell^e}
\newcommand{\lnu}{\ell^\nu}
\newcommand{\lnuc}{\ell^{\nu c}}
\newcommand{\ble}{\bar{\ell}^e}
\newcommand{\blnu}{\bar{\ell}^\nu}

\newcommand{\lte}{\tilde \ell^e}
\newcommand{\blte}{\bar{\tilde \ell}^e}

\newcommand{\ltnu}{\tilde \ell^\nu}
\newcommand{\ltnuc}{\tilde \ell^{\nu c}}
\newcommand{\bltnu}{\bar{\tilde \ell}^\nu}

\newcommand{\et}{\tilde e} 
\newcommand{\bet}{\bar{\tilde e}} 
\newcommand{\dslash}{\not \!\partial} 
\newcommand{\nova}[1]{}

\begin{document}

\topmargin -0.5cm

\preprint{UdeM-GPP-TH-09-183}
\preprint{IFIBA-TH-09-001}
 
\title{Vertex Displacements for Acausal Particles: Testing the
Lee-Wick Standard Model at the LHC}

\author{Ezequiel \'Alvarez} 
\affiliation{CONICET, IFIBA and Departamento de
F\'{\i}sica, FCEyN, Universidad de Buenos Aires, Ciudad Universitaria,
Pab.1, (1428) Buenos Aires, Argentina} 

\author{Leandro Da Rold} \affiliation{CONICET and Centro At\'omico
Bariloche, Av.\ Bustillo 9500, 8400, S.\ C.\ de Bariloche, Argentina }

\author{Carlos Schat} 
\affiliation{CONICET, IFIBA and Departamento de
F\'{\i}sica, FCEyN, Universidad de Buenos Aires, Ciudad Universitaria,
Pab.1, (1428) Buenos Aires, Argentina} 
\affiliation{Department of
Physics and Astronomy, Ohio University, Athens, Ohio 45701, USA}

\author{Alejandro Szynkman} 
\affiliation{Physique des Particules, Universit\'e de
Montr\'eal, C.P. 6128, succ.\ centre-ville, Montr\'eal, QC,
Canada H3C 3J7}

\date{\today}

\begin{abstract} 
We propose to search for wrong displaced vertices, where decay
products of the secondary vertex move towards the primary vertex
instead of away from it, as a signature for microscopic violation of
causality.  We analyze in detail the leptonic sector of the recently
proposed Lee-Wick Standard Model, which provides a well motivated
framework to study acausal effects.  We find that, assuming Minimal
Flavor Violation, the Lee-Wick partners of the electron, $\lte$ and
$\et$, can produce measurable wrong vertices at the LHC, the most
promising channel being  $q \bar q \longrightarrow \blte \lte
\longrightarrow e^+ e^- jjjj$.  A Monte-Carlo simulation using
MadGraph/MadEvent suggests that for $M_\ell \lesssim 450 \
{\rm GeV} $ the measurement of these acausal vertex displacements
should be accessible in the LHC era.
\end{abstract}

\maketitle

\section{Introduction}

Despite all the phenomenological success of the Standard Model (SM),
the large quantum corrections to the Higgs potential require a fine
tuning that makes it unnatural as a complete theory. This fine tuning
is known as the hierarchy problem and can be avoided if there is new
physics at the TeV scale, the energy scale that will be explored at
the Large Hadron Collider (LHC). Finding a solution to the hierarchy
problem that can also be tested soon by the upcoming experiments at
the LHC provides a very strong motivation for building extensions to
the Standard Model.

A recent proposal that solves the hierarchy problem is the Lee-Wick
Standard Model (LWSM)~\cite{Grinstein:2007mp}, based on ideas of Lee
and Wick~\cite{Lee:1969fy,Lee:1970iw} for a finite theory of quantum
electrodynamics.  In the LWSM each SM particle has a Lee-Wick (LW)
partner of the same statistics. The only new parameters of the model
are the LW mass matrices.  LW particles have a kinetic term with the
opposite sign to the usual one for SM particles, leading to partial
cancellations in loop corrections that eliminate quadratic
divergences.

Different theoretical and phenomenological aspects of the LWSM have
been discussed recently in Refs.~\cite{Grinstein:2007iz,
Fodor:2007fn,Knechtli:2007ea,Grinstein:2008qq, FCNC, Wu:2007yd,
Espinosa:2007ny, Rizzo:2007ae, Krauss:2007bz, Carone:2008bs, Rizzo:2007nf,
Carone:2008iw, Carone:2009it, ewpt, Underwood:2008cr, O(N), Fornal:2009xc}.

A very interesting and distinctive feature of the LWSM is its acausal
behavior. This might be a serious problem of the theory if acausal
effects could persist to macroscopic scales, leading to paradoxes. For
a general discussion of causality see Coleman lectures,
Ref.~\cite{coleman}. It has been argued that if LW particles decay
fast enough the violation of causality would happen on a very small
time scale and macroscopic causality would still be preserved as an
emergent property~\cite{O(N)}.

Besides its phenomenological interest, LW-type theories also give the
theoretical framework to discuss acausality. A LW version of the O(N)
model has been used to check the consistency of the acausal theory to
all orders in perturbation theory~\cite{O(N)}.  The authors have shown
that there is a well-defined S-matrix in scattering processes, that
provides a one-to-one map from the past to the future.  Although a
similar result for the LWSM would be much more difficult to obtain,
the result of Ref.~\cite{O(N)} is encouraging.  The properties in
thermal equilibrium have also been examined to further check if
multiple scattering can lead to macroscopic acausal
behavior~\cite{Fornal:2009xc}.

The main question we address here is the following. Given a theory
that allows for microscopic violations of causality, but which is
still free from paradoxes at the macroscopic level: Is it possible to
propose an observable that could reveal a microscopic violation of
causality by solely analyzing the {\it in} and {\it out} states of the
S-matrix?  We answer this question in the affirmative and to this
purpose we define a {\it wrong vertex displacement} as a vertex
displacement in which the decay products coming from the secondary
vertex have a total momentum that points from the secondary to the
primary vertex and its invariant mass corresponds to a new
resonance\footnote{A similar experimental signature, but without
requiring the total momentum of the decay products to have a fixed
invariant mass, is called ``large negative impact parameter (LNIP)''
in Ref.~\cite{lnip}.}. As we discuss in the next Section, acausal
theories give rise to wrong vertex displacements.
 
As a direct application of the general question here addressed and the
above defined observable, we analyze in this article the possibility
of measuring a wrong vertex displacement as a signature of a LW
resonance at the LHC.  We investigate which LW particles could have
the smallest widths to produce the largest wrong vertex displacements.

To answer this question we study in detail the flavor interactions of
the leptonic sector in the LWSM. We obtain that if the LW sector
satisfies Minimal Flavor Violation (MFV), the best candidate to
produce wrong displaced vertices is the LW electron associated to the
SU(2) doublet. In order to investigate the possibility of measuring a
wrong vertex displacement at the LHC, we perform Monte Carlo
simulations using MadGraph/MadEvent \cite{mgme} to study which are the
conditions such that there is a LW particle stable enough to produce
an observable wrong vertex displacement.  We find that for LW masses
satisfying $M_{\ell} \lesssim450\ {\rm GeV} $, these particles would
produce displacements in the transverse plane greater than $20\,{\rm
\mu m}$ at cross-sections that would be measurable in the LHC era.
Another good candidate to produce observable wrong vertex
displacements is the LW partner of the electron in the singlet
representation of the gauge group.  Other LW particles seem to be out
of the reach of the LHC for these purposes, but could still be
observed by direct production, although it would be more difficult to
discriminate them from other candidates of new physics. A wrong
displaced vertex on the other hand, would be a characteristic
signature for an acausal particle.

The paper is organized as follows. In Section \ref{sec-wvd} we show
that acausal resonances lead to wrong vertex displacements and argue
that, in the LWSM, LW leptons are the best candidates to observe them.
In Section \ref{sec-flavor} we analyze the flavor structure in the
leptonic sector of the LWSM, and we obtain the mass eigenstates and
the interactions in this basis. In Section \ref{sec-stable} we compute
the widths for different LW-leptonic physical states and conclude that
the first generation of LW-charged leptons will have a considerably
smaller width than other LW particles. In Section \ref{sec-experiment}
we discuss which are the necessary conditions to identify the desired
events and we design a set of cuts and procedures to isolate our
signal at the LHC.  We end with our conclusions in Section
\ref{sec-conclusions}.  Details of the diagonalization of the leptonic
mass matrix and the interactions between mass eigenstates can be found
in the Appendices.

\section{Wrong vertex displacements}\label{sec-wvd}

One of the main purposes of this article is to propose the measurement
of wrong vertex displacement associated to processes that go through
LW resonances at the LHC. A normal (non-LW) resonant state, produced
by some initial particle collision, propagates typically for the space
of time allowed by its mean lifetime before it decays into lighter
particles. Unlike the usual case, the opposite occurs for LW
resonances: decay products precede production. However, for the theory to
be consistent it should forbid the temporal exploration of this
time-scale.  On the other hand, it is possible to probe spatially this
acausal behavior by means of the detection of displaced decay vertices
corresponding to relatively `long-lived' LW resonances and through the
measurement of certain kinetic variables related to the decay
products.

When a normal resonance is created with non-zero momentum, the total
momentum of the decay products points from the decay point (secondary
vertex), in the outgoing direction from where the resonance was
created (primary vertex)\footnote{A familiar case where this has been
observed corresponds to the $B$ mesons. After these particles are
created, they travel some distance in the laboratory frame before they
decay. One can infer the time elapsed between the production and decay
points to obtain a measure of the $B$ mesons mean lifetimes. This kind
of measurements have been performed by using different techniques at
LEP, Tevatron and the $B$ factories~\cite{blifetime}.}. Of course,
this is a direct consequence of the outgoing direction of the momentum
of the resonance.

On the other hand, the presence of LW --or any acausal-- resonances
may be detected by observing an opposite pattern: the resulting
momentum of the decay products heads inwards, from the secondary
towards the primary vertex.

\begin{figure}[t]
\begin{center}
\includegraphics[width=.46\textwidth]{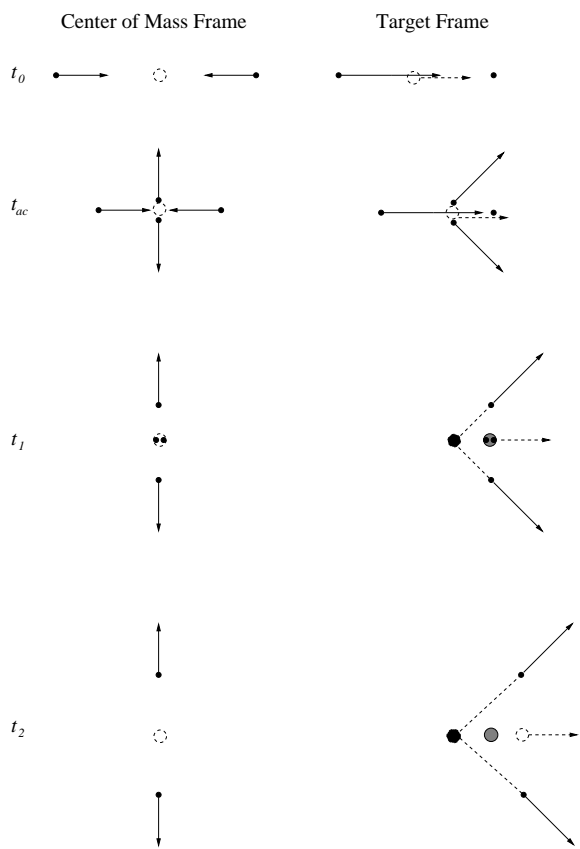} 
\caption{Pictorial description of the collision and decay process
through an intermediate LW resonance, as seen in two reference
frames. The dashed circle is the center of mass of the system. The
gray (black) disk signals the position of the primary (secondary)
vertex in the target frame.  The acausal behaviour sets in at time
$t_{ac}$, when the decay products appear before the LW resonance is
created. At long times before and after the collision ($t_0, t_2$) the
configuration seems equivalent to the causal case, except that the
extrapolated secondary vertex lies to the {\it left} of the primary
vertex and the decay products move {\it towards} the primary vertex
instead of away from it. As it is explained in the text, the same
final result ($t_2$ as an {\it out} state of a $t_0$ {\it in} state)
is rigorously derived using the S-matrix formalism without inquiring
what happens {\it during} the process, that is, without a microscopic
description between $t_{ac}$ and $t_1$.}
\label{fig:acausal} 
\end{center} 
\end{figure}

This particular behavior may be understood by means of a quantum
mechanical argument as well as in a formal way using an S-matrix
description. In the first case, Lee and Wick
~\cite{Lee:1969fy,Lee:1970iw} and Coleman ~\cite{coleman} have argued
that, in a process going through a LW resonance in the center of mass
frame, decay products would appear {\it before} collision takes
place. In this picture, incoming particles collide in the same spatial
point where decay products come from.  Moreover, incoming and outgoing
particles have zero total momentum in this frame of reference. If this
whole process is boosted in some direction and its events are
accordingly Lorentz transformed, one easily retrieves the wrong
displaced vertex described in the previous paragraph.  (Moreover, just
a Galileo transformation is enough for this purpose.) We show
pictorially in Fig.~\ref{fig:acausal} this production and decay
process mediated by an acausal resonance in both reference frames.

Alternatively, and in a more formal way, a recent paper by Grinstein,
O'Connell and Wise~\cite{O(N)} may also help us to understand this
behavior. They compute the transition amplitude between arbitrary
two-particle states mediated by LW resonances using the S-matrix
description of quantum field theory. The final result of this
calculation, which corresponds to Eq. (46) in their paper, is the
following

\bea 
\langle \psi_{\tiny{\mbox{out}}} | \psi_{\tiny{\mbox{in}}} \rangle
\simeq \frac{g^2 \sqrt{M}}{2 {(2 \pi \sqrt{\omega^2})}^{3/2}}
\hat{F}(-M \omega / \sqrt{\omega^2}) \hat{G}(-M \omega /
\sqrt{\omega^2}) e^{i M \sqrt{\omega^2}} e^{-\Gamma \sqrt{\omega^2}
/2} \, .
\label{amplitude} 
\eea 
The relevant argument for our analysis lies in the functions $\hat{F}$
and $\hat{G}$ which essentially contain the information about the
distribution of momenta in the initial and final states,
respectively. Both $\hat{F}$ and $\hat{G}$ are peaked
around\footnote{Instead, the relation $q \approx M \omega /
\sqrt{\omega^2}$ is satisfied by a causal resonances. See Eq. (34) in
Ref.~\cite{O(N)}.} $q \approx - M \omega / \sqrt{\omega^2}$, where $q$
stands for the total four-momentum of the incoming/outgoing particles,
$\omega = z_D - z_P$ is the space-time separation between the
positions of the decay ($z_D$) and production ($z_P$) vertices
associated to the intermediate LW-resonant state, and $M$ represents
its mass. Due to the relative sign between $q$ and $\omega$, the
secondary vertex is displaced from the primary towards the opposite
direction of $q$. Consider now the case where the total momentum of
the incoming particles is different from zero in the laboratory
reference frame, as for instance $t_0$ in the target frame in
Fig.~\ref{fig:acausal}.  This means that if $q$ points towards the
right in the figure, then the traces of the decay products should
appear to come from a secondary vertex located at the {\it left} of
the primary vertex, as it happens at $t_2$ in the target frame in
Fig.~\ref{fig:acausal}.  As a matter of fact in an S-matrix
description one has access only to these two times ($t_0$ and $t_2$),
since it is senseless to inquire what happens {\it during} the
process. This is the way the atypical displacement pattern of
secondary vertices arises in scattering processes mediated by LW
resonances in the S-matrix description. Other consequences of
Eq.~(\ref{amplitude}) are further discussed in Ref.~\cite{O(N)}. As a
last remark, we mention that this equation has been derived in the
narrow resonance approximation which is valid in the case under study
here (typically, for a LW electron with M $\sim$ 300 GeV, we find
$\Gamma \sim 10^{-11}$ GeV).

There is a point we should clarify in order to avoid misunderstandings
in the interpretation of what we have discussed in the previous
paragraphs. Within the context of the S-matrix formalism, it does not
make sense to rise the question about the time nor the position the
vertices take place since we only deal with asymptotic states in this
framework. In order to establish a proper connection between this
theoretical approach and observations, it is necessary to perform
measurements far away from the region where the resonance propagates
and at times well separated from the interval of time the propagation
takes. These conditions are indeed satisfied at the LHC: measurements
are carried out far outside the interaction region and within a
relative long-time window that contains the entire process of creation
and decay of the LW resonances. Thus, the vertex positions are not
measured (in the sense that measurements do not perturb the dynamics
of the system at the vertices) but only indirectly obtained by
extrapolating the traces left by the outgoing particles along the
detectors.

Once we have justified how the LWSM (or any theory with acausal
particles) could give rise to a wrong vertex displacement, we may
focus on how this signature could be observed in this specific model.

The first important question to address is about the existence of
general reasons to expect a LW particle to have a very small width.
The width is determined by the masses and the interactions between the
particles involved in the decay process.  Present constrains on the LW
masses coming from EWPT~\cite{ewpt,Carone:2008iw,Underwood:2008cr}
indicate that the lowest allowed masses for LW quarks and
LW-intermediate bosons are close to 3 TeV, whereas the Higgs sector is
only constrained to have masses above $\sim 400$ GeV \cite{new}. On
the the other hand, there are no limits on LW-leptonic masses other
than those coming from direct lepton search, which give a lowest bound
of $\sim 100$ GeV \cite{PDG}.
Besides the mass range difference between the LW leptons and the LW
quarks, these last have the additional interaction with the gluon and
the LW gluon.  Therefore, the LW quarks are expected to have a total
width greater than the LW leptons.  A similar reasoning holds for the
LW-intermediate bosons, which have stronger constraints and prompt
decays. Hence, we conclude that LW leptons may be expected to be less
unstable\footnote{A few days before the submission of this work,
Ref.\cite{new} reported a low bound for the Higgs mass in $\sim400$
GeV. Although we have not performed the due analysis, we expect that
the many decay channels open for the LW Higgs would increase its width
considerably above the one expected for the LW leptons. In any case,
we clarify that this does not modify our results.} and, therefore,
the more propitious to give a wrong vertex displacement signature.

\begin{figure}[h] 
\includegraphics[width=.6\textwidth]{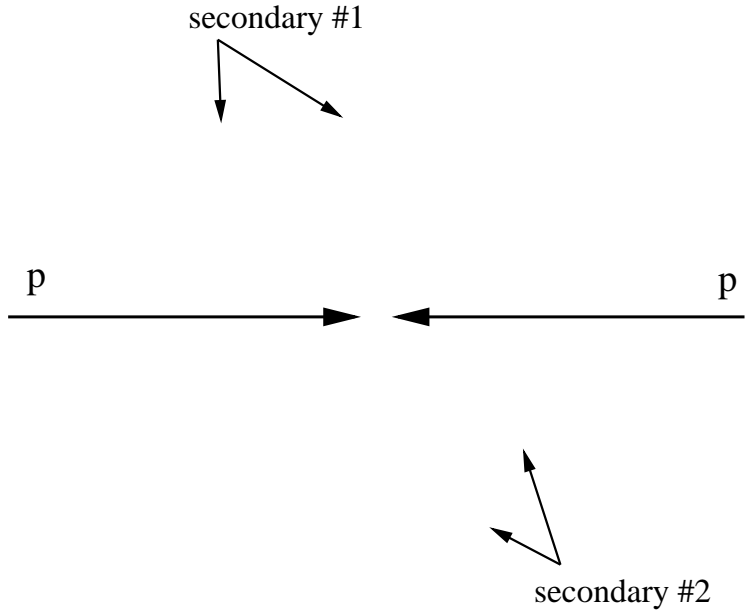}
\caption{Two wrong vertex displacement expected in the LWSM. Decay
products coming from both secondary vertices travel towards the
primary vertex, instead of moving away from it.}
\label{fig:4}
\end{figure}

The next important feature when explicitly analyzing the LWSM is that,
since the creation of a single LW lepton $\lte$ is suppressed in this
model, is more likely to observe two wrong displaced vertices than
one. For instance in the process $pp \to \blte \lte \to X_1 X_2$ both
$\lte$'s would produce wrong vertex displacements and hence their
decay products $X_1$ and $X_2$ would have their respective total
momentum pointing towards the primary vertex.  This situation, as
expected at the LHC, is shown in Fig.~\ref{fig:4}.

In the next Section we study the LW-leptonic sector in order to obtain
the mass eigenstates that should be created to give the sought signal,
and their interactions.

\section{Flavor in the leptonic sector of the LWSM}\label{sec-flavor}

The LWSM is usually formulated in terms of flavor eigenstates.  The
particle content of the fermion fields of the model is shown in
Table~\ref{table-particle-content}. A generation index is omitted,
however when along the text it is required to specify the generation,
a corresponding sub-index will be added.

The Yukawa interactions mix the SM fields with the LW fields and are not
diagonal in the generation space. In order to understand flavor issues
in the LWSM, and to be able to determine which are the `longest-lived'
acausal states, it is suitable to work in the mass eigenstates basis.
With this objective, in this Section we first show how a change of
basis may isolate the Yukawa terms within each generation, and then
how a second change of basis performed within each generation gives
the final physical states. We finish with an analysis of the
interactions resulting from these physical states .

\begin{table}[h] 
\begin{center} 
\begin{tabular}{ccc} 
\hline\hline leptons
& quarks & SU(2) \\ \hline & & \\ $ \left( \begin{array}{c} \lnu 
\\ \le    \end{array} \right)_L $ \,, $ \left( \begin{array}{c}
\ltnu  \\ \lte    \end{array} \right)_{L,R} $
\qquad \qquad & $ \left( \begin{array}{c} q^u  \\ q^d    \end{array}
\right)_L $ \,, $ \left( \begin{array}{c} \tilde q^u  \\ \tilde q^d
\end{array} \right)_{L,R} $ & \bf{2} \\[.2in] $\nu_R $  &  $ u_R \,,
\tilde u_{L,R}$ & \bf{1} \\[.1in] $e_R$ \,, $\et_{L,R}$ & $ d_R \,,
\tilde d_{L,R}$ & \bf{1}
\\[.1in] \hline\hline 
\end{tabular} 
\end{center} 
\caption{Fermionic fields in the LWSM.  The neutrinos are Majorana
particles. The right handed neutrino $\nu_R$ does not have a LW
partner. The dimension of the SU(2) weak isospin representation is
indicated in the last column.  A generation subindex is omitted here,
but it will be shown whenever necessary.}
\label{table-particle-content}
\end{table}

\subsection{Mixing between generations} After the Higgs breaks the EW
symmetry, the mass terms for the charged leptons in the LWSM are given
by: 
\begin{equation}
\label{Le-flavor} 
{\cal L}_e=-(\bar e_R-\bar \et_R)m'_e(\le_L-\lte_L)+\blte_R M_\ell \lte_L + \bar \et_R M_e \et_L + \rm{h.c.} , 
\end{equation} 
\noindent where $M_{\ell,e}$ are the mass matrices of the LW lepton doublet and singlet, 
and $m'_e=y_e v/\sqrt{2}$ is a $3\times3$ matrix in generation space, where generation
indices are understood. Notice, for future purposes, that although the LW leptons are 
vector-like, only the Left- (Right-) chirality of the LW-doublet (singlet) is present
in the the Yukawa interactions.

The Yukawa mixings between different generations can be diagonalized
performing the following SM rotations:
\begin{eqnarray}
\label{rotation-e} 
e_R\rightarrow A^e_R e_R \ , \qquad
\et\rightarrow A^e_R \et \ ,\nonumber \\ \le_L\rightarrow A^e_L
\le_L \ , \qquad \lte\rightarrow A^e_L \lte \ ; 
\end{eqnarray} 
where $A^e_{L,R}$ are the usual unitary matrices of the SM
diagonalizing the Yukawa couplings of the charged leptons. Note that
we have transformed both the Left- and Right-handed components of $\et$
($\lte$) with the same matrix $A^e_R$ ($A^e_L$). In this new basis the
mass terms become: 
\begin{equation}
\label{Le-flavor-diagonal} 
{\cal L}_e=-(\bar e_R-\bar \et_R)m_e(\le_L-\lte_L)+\blte_R A^{e\dagger}_L
M_\ell A^e_L\lte_L + \bar \et_R A^{e\dagger}_R M_e A^e_R \et_L +
\rm{h.c.} , 
\end{equation} 
\noindent with $m_e$ the diagonal mass
matrix of the SM. If we impose MFV in the LW sector, the matrices
$M_{\ell,e}$ are proportional to the identity and we can trade
$M_{\ell,e}\rightarrow M_{\ell,e}{\bf 1}_{3\times3}$, with
$M_{\ell,e}$ ordinary numbers. Therefore,
Eq.~(\ref{Le-flavor-diagonal}) is diagonal in the generation space and
one can consider the mixings between SM and LW leptons for each
generation isolated from the others, as far as the mass terms are
concerned\footnote{If we relax MFV and allow for general
$M_{\ell,e}$, it is not possible to diagonalize the mass matrix in the
generation space with the usual SM transformation, and the mixings
depend on the unknown parameters of the LW sector, see
Ref.~\cite{FCNC}.}.

We consider now the neutral leptons of the LWSM.
Ref.~\cite{Espinosa:2007ny} showed that it is possible to preserve the
familiar see-saw mechanism, without destabilizing the Higgs mass, by
introducing a very heavy Right-handed neutrino, and no LW partner is
needed for $\nu_R$. The corresponding mass Lagrangian is:
\begin{equation}
\label{Lnu-flavor} 
{\cal L}_\nu=-\bar \nu_R m'_\nu(\lnu_L-\ltnu_L)-\frac{1}{2}\bar\nu^c_R m_R\nu_R+\bltnu_R M_\ell
\ltnu_L + \rm{h.c.} , 
\end{equation} 
\noindent with $m_R$ the Majorana
mass of the Right-handed neutrino and a non-diagonal Dirac mass
$m'_\nu=y_\nu v/\sqrt{2}$. As before, the mixings between generations
can be diagonalized with the usual SM transformations:
\begin{eqnarray}
\label{rotation-nu} 
\nu_R\rightarrow A^\nu_R \nu_R \ ,
\qquad \lnu_L\rightarrow A^\nu_L \lnu_L \ , 
\qquad \ltnu\rightarrow
A^\nu_L \ltnu \ .  
\end{eqnarray} 
\noindent Note again that both
chiralities of $\ltnu$ transform in the same way. Assuming MFV in the
LW sector, the LW mass term remains diagonal. The only possible source
of flavor mixing in the mass terms is the Majorana mass, and thus the
generation mixings are suppressed by this very high scale.

We consider now the effect of the previous transformations in the
interactions. Since the neutral current interactions are invariant
under those transformations, the interactions with the photon, the
$Z$, the neutral Higgs and their LW partners do not change flavor.
Particularly important for the phenomenology are the following terms:
\begin{eqnarray}
\label{LNC} 
\mathcal{L}_{NC}=-(Z_\mu+\tilde Z_\mu)
\large[g_z^{e_L}(\ble_L\gamma^\mu
\le_L-\blte\gamma^\mu\lte)+g_z^{e_R}(\bar e_R\gamma^\mu
e_R-\bet\gamma^\mu \et) + g_z^{\nu_L}(\blnu_L\gamma^\mu
\lnu_L-\bltnu\gamma^\mu\ltnu)\large] \ , \ \ \\
\mathcal{L}_{NY}=-\frac{y_e}{\sqrt{2}}(\bar e_R-\bet_R)(h-\tilde
h+i\tilde P)(\le_L-\lte_L) -\frac{y_\nu}{\sqrt{2}}\bar \nu_R
(h-\tilde h-i\tilde P)(\lnu_L-\ltnu_L)+ \rm{h.c.}\ , \ \ 
\label{LNY}
\end{eqnarray} 
\noindent where $g_z^{e_{L,R}}$ and $g_z^{\nu_L}$ are
the $Z$ couplings of the SM leptons, $y_{e,\nu}$ are the diagonal
Yukawa couplings, and $\tilde h$ $(\tilde P)$ stands for the neutral 
scalar (pseudoscalar) component of the LW-Higgs field.

As usual, the charged current (CC) interactions are flavor changing. In
terms of the new fields we obtain: 
\begin{eqnarray}
\label{LCC}
\mathcal{L}_{CC}=-\frac{g_2}{\sqrt{2}}(W^+_\mu+\tilde W^+_\mu) (
\blnu_L\gamma^\mu V \ell_L-\bltnu\gamma^\mu V \lte) + \rm{h.c.}\ ,\\
\mathcal{L}_{CY}=y_e(\bar e_R-\bet_R)V^\dagger \tilde h^-
(\lnu_L-\ltnu_L) -y_\nu\bar \nu_R V\tilde h^+ (\le_L-\lte_L)+
\rm{h.c.}\ ,\label{LCY} 
\end{eqnarray} 
\noindent where $V=A^{\nu\dagger}_L A^e_L$ is the usual
Pontecorvo-Maki-Nakagawa-Sakata leptonic mixing
matrix~\cite{VPMNS}. We can see from Eqs.~(\ref{LCC}) and~(\ref{LCY})
that, assuming MFV, all the flavor changing interactions are
determined by the SM parameters.

\subsection{Mixings between SM and LW leptons} 
The charged lepton mass
eigenstates can be obtained by making a simplectic rotation for each
generation, similar to the one performed in Ref.~\cite{ewpt, Underwood:2008cr}. For each generation, we define three
dimensional flavor eigenvectors containing the SM charged lepton and
their LW partners, $\et$ and $\lte$: 
\begin{equation}
\label{defE1}
E_L^t=(\le_L,\et_L,\lte_L) \ , \qquad E_R^t=(e_R,\et_R,\lte_R) \ ,
\end{equation} 
\noindent (note that we have changed the basis order
compared with Refs.~\cite{ewpt,Underwood:2008cr}) and three
dimensional mass eigenstate vectors: \begin{equation}\label{defE2}
\mathcal{E}_L^t=(\mathcal{E}^1_L,\mathcal{E}^2_L,\mathcal{E}^3_L) \ ,
\qquad
\mathcal{E}_R^t=(\mathcal{E}^1_R,\mathcal{E}^2_R,\mathcal{E}^3_R) \ ,
\end{equation} related by a simplectic rotation in the following way:
\begin{equation}\label{rotationE} E_{L,R}=S^e_{L,R}\mathcal{E}_{L,R} \
.  \end{equation} \noindent Expanding in powers of Yukawa insertions,
at leading order the mass eigenstates coincide with the flavor
eigenstates, and the mixings are suppressed by powers of
$m_e/M_{\ell,e}$, 
\begin{equation} 
\mathcal{E}_{L,R} = E_{L,R} + {\cal O} (m_e/M_{\ell,e});
\end{equation} 
where $\mathcal{E}^1_{e,\mu,\tau}$ are the usual electron, muon and tau.
We show the diagonalization using
this approximation in the Appendix~\ref{Ap-diagonalization}. 
 
The neutral lepton mass eigenstates are Majorana fermions, and all the
mixings are suppressed by at least one power of $m_R$. Similarly to
the charged leptons, we define a vector containing four Majorana
neutrinos for each generation: 
\begin{equation}
\label{defN1}
N^t=(\lnu_L+\lnuc_L,\nu_R+\nu_R^c,\ltnu_L+\ltnuc_L,\ltnu_R+\ltnuc_R)
\ , 
\end{equation} 
\noindent and a four dimensional mass eigenstate
vector: 
\begin{equation}
\label{defN2}
\mathcal{N}^t=(\mathcal{N}^1,\mathcal{N}^2,\mathcal{N}^3,\mathcal{N}^4)
\ , \end{equation} related by a simplectic rotation:
\begin{equation}
\label{rotationN} 
N=S^\nu\mathcal{N} \ .
\end{equation} 
\noindent There is a light neutrino $\mathcal{N}^1$, whose mass is
given, at leading order in Yukawa insertions, by the usual see-saw
mechanism. There are two neutrinos $\mathcal{N}^{3,4}$ that can be
associated with the degrees of freedom of $\ltnu$, with masses
$M_\ell+\mathcal{O}(m_\nu)$, and a fourth heavy neutrino
$\mathcal{N}^2$ that can be associated with $\nu_R$, up to corrections
of order $\sim {\cal O}(\frac{m_\nu}{m_R})$. We show the details of
the diagonalization in the Appendix~\ref{Ap-diagonalization}.  As an
interesting aspect of this result, we notice that the usual see-saw
mechanism is not destabilized by the addition of a LW neutrino with
$\sim$ TeV mass.

\subsection{Interaction features of the mass eigenstates}

To obtain the interactions between the mass eigenstates one has to
perform the above simplectic rotations of the charged and neutral
leptons in the interaction terms. This can be done immediately by
using the rotation matrices of Appendix~\ref{Ap-diagonalization}. The
details of this calculation are found in
Appendix~\ref{Ap-interactions}, where we explicitly show the
interactions between the mass eigenstates.  

Once we have the interactions of the mass eigenstates its decay
properties are analyzed straightforward.  For the purposes we follow,
we are interested in the decay of the acausal charged and neutral
leptons $\mathcal{E}^{2,3}$ and $\mathcal{N}^{3,4}$.  

We begin analyzing neutral currents (NC). These interactions do not
change flavor, however 
they can produce the interaction between different mass eigenstates.
This is not the case for the electromagnetic interaction, since the SM
leptons and their LW partners have the same electromagnetic charge and
hence heavy leptons can not decay at tree level electromagnetically.
On the other hand, the interactions with $Z$ (and $\tilde W^3$ and
$\tilde B$) mix different mass eigenstates within a given generation
because not all the leptons with a given chirality have the same
$Z$-coupling, for example: $\et_L$, and $\le_L$ and $\lte_L$, although being
all Left-handed charged fermions --see Eq.~(\ref{defE1})-- have different
$Z$-couplings. Therefore, the heavy charged leptons can decay via
${\cal E}_{2,3} \to Z+\mathcal{E}^1$, with a suppression factor
$m_e/M_{\ell,e}$ in the amplitude.  In the flavor basis this is
understood as a Yukawa insertion times a suppressing LW-fermionic
propagator, as shown in Fig.~\ref{fig-decays-et}. The heavy neutral
leptons can decay by a similar process to $Z+\mathcal{N}^1$, but in this case
with an extra Majorana suppression $m_\nu/m_R$.

The neutral Higgs interactions (NY) mix mass eigenstates. The charged
leptons $\mathcal{E}^{2,3}$ interact with the light leptons
$\mathcal{E}^1$ with Yukawa strength $y_e$ and no extra suppression.
The coupling between neutral leptons $\mathcal{N}^{3,4}$ and
$\mathcal{N}^1$ is proportional to $y_\nu$, but it is suppressed by
$m_\nu/m_R$ since a virtual $\nu_R$ is needed to generate this
interaction.

The charged current interactions also mix mass eigenstates and in
addition, as usual, different generations. The heavy charged leptons
${\cal E}^{2,3}$ can decay to $W+\mathcal{N}^1$, with a coupling
constant proportional to the leptonic mixing matrix
$V$.
The dominant charged current decays have amplitudes suppressed by
$m_e/M_e$ for ${\cal E}^2$ and by $m_e^2/M_{\ell,e}^2$ for ${\cal
E}^3$, where the Yukawa insertion $m_e$ has the same generation index
as the decaying lepton. It is interesting to notice that this
different suppression factor could also be understood through the
flavor basis point of view, where the key difference comes from the
Yukawa couplings, since the Yukawa terms only couple $\lte_L$ to
$\et_R$. On the other hand, the amplitudes for the neutral leptons
decay, $\mathcal{N}^{3,4} \to \mathcal{E}^1 + W$, is proportional to
the corresponding $V$ and is suppressed by $m_e/M_e$, but the Yukawa
insertion $m_e$ has the generation index corresponding to the final
lepton $\mathcal{E}^1$.  This is a crucial difference, since the
contribution from a tau as a final lepton enhances the LW-neutrinos
width.

The interactions with the charged LW Higgs $\tilde h^\pm$ (CY) mix
mass eigenstates and different generations. The coupling which drives
${\cal E}^{2,3} \to\tilde h ^\pm {\cal N}^1 $ is proportional to $y_e$
and $V$, but the one corresponding to $\mathcal{E}^3$ has an extra
suppression $\sim m_e/M_{\ell,e}$. Again, this difference is traceable
to the different chiralities of the LW fields entering into the Yukawa
interactions. At leading order, the coupling in charge of
$\mathcal{N}^{3,4} \to \tilde h ^\pm \mathcal{E}^1$ is proportional to
the Yukawa of the final charged lepton, $y_e$, without extra
suppressions. Again, the tau contribution will enhance the
LW-neutrinos width through this channel.

Using Appendix \ref{Ap-interactions}, we summarize the relevant interactions for the decay of the LW-mass
eigenstates ${\cal E}^{2,3}$ and ${\cal N}^{3,4}$ in Table
\ref{table-interactions}.  
\begin{table}[h] 
\begin{center}
  \begin{tabular}{| c | c | c | c | c |} 
\hline lepton$|$interaction  & NC & NY & CC & CY \\ \hline 
    $\mathcal{E}^2_\beta$ &
$\frac{m_e^\beta}{M_e}\delta^{\alpha\beta}$ & $y_e^\beta
\delta^{\alpha\beta}$ &  $\frac{m_e^\beta}{M_e}V^{a\beta}$ &
$y_e^\beta V^{a\beta}$ \\ \hline 
    $\mathcal{E}^3_\beta$ & $\frac{m_e^\beta}{M_\ell}\delta^{\alpha\beta}$ &
$y_e^\beta \delta^{\alpha\beta}$ &
$\left(\frac{m_e^\beta}{M_{e,\ell}}\right)^2V^{a\beta}$ & $y_e^\beta
\frac{m_e^\beta}{M_{e,\ell}}V^{a\beta}$ \\ \hline 
    $\mathcal{N}^{3,4}_b$ & $\frac{(m_\nu^b)^2}{M_\ell
m_R}\delta^{ab}$ &
$y_\nu^a\left(\frac{m_\nu^a}{m_R}\mp\frac{m_\nu^a}{M_\ell\mp
m_R}\right) \delta^{ab}$ &  $\frac{m_e^\alpha}{M_\ell}V^{\dagger \
\alpha b}$ & $y_e^\alpha V^{\dagger \ \alpha b}$ \\ \hline 
  \end{tabular} \end{center} \caption{Relevant interactions for the
decay of the LW leptons. We show explicitly the indices $a,b=1,2,3$
that number the generations of neutrino mass eigenstates, and
$\alpha,\beta=e,\mu,\tau$ that correspond to flavor. $m^\alpha_e$ and
$m^a_\nu$ stand for the corresponding Dirac masses:
$m^\alpha_e=y^\alpha_e v/\sqrt{2}$ and $m^a_\nu=y^a_\nu v/\sqrt{2}$.
For the NC interactions involving $\mathcal{E}^i$ or $\mathcal{N}^i$
and $Z$ we show only the decaying lepton in the first column, thus the
couplings correspond to $\mathcal{E}^{2}_\beta\to
Z+\mathcal{E}^1_\alpha,\mathcal{E}^{3}_\beta\to
Z+\mathcal{E}^1_\alpha$ and $\mathcal{N}^{3,4}_b\to
Z+\mathcal{N}^1_a$, for the first, second and third line respectively.
A similar situation holds for the other interactions that drive the
following decays: NY($\mathcal{E}^{2,3}_\beta\to
h+\mathcal{E}^1_\alpha,\mathcal{N}^{3,4}_b\to h+\mathcal{N}^1_a$),
CC($\mathcal{E}^{2,3}_\beta\to
W+\mathcal{N}^1_a,\mathcal{N}^{3,4}_b\to W+\mathcal{E}^1_\alpha)$,
CY($\mathcal{E}^{2,3}_\beta\to \tilde
h+\mathcal{N}^1_a,\mathcal{N}^{3,4}_b\to \tilde
h+\mathcal{E}^1_\alpha$).  This table is obtained from Appendix \ref{Ap-interactions} results.} 
\label{table-interactions} 
\end{table}

\section{Width of Narrow Lee-Wick resonances}\label{sec-stable}

In this Section we explicitly compute the width of narrow LW
resonances.  As we will see, the magnitude of these widths will select
the first generation ${\cal E}^2_{e}$ and ${\cal E}^3_{e}$ --the
subindex indicates the first generation-- as the best candidates to
produce wrong vertex displacements at the LHC.  It is worth to mention
that this study is based on the analysis of two-body decays, unless
something different is stated.

In the previous Section we found that ${\cal E}^{2,3}_\alpha$ decays
were suppressed by the Yukawa $y_e^\alpha$, that corresponds to the
generation of the decaying LW particle.  Therefore, as a general
feature, the mass states associated to the first generation are more
stable than the others.  On the other hand, the dominant decays of
${\cal N}^{3,4}$ involve charged interactions that mix generations,
and are proportional to the Yukawa of the final charged lepton. Hence,
the tau channel dominates and gives a larger width for these
LW neutrinos.  We compute explicitly the decay width for the ${\cal
E}^{2,3}$ and ${\cal N}^{3,4}$ LW-mass eigenstates of the first
generation.

\subsection{$\mathcal{E}^2$ total width}
The NC decay $\mathcal{E}^2\to Z+\mathcal{E}^1$ is suppressed by
$m_e^2/M_e^2$ (the partial width is proportional to the coupling
square), without flavor change. 
A similar suppression factor is present for the decay through CC
interactions, but in this case the final neutrino can have any flavor,
with a coefficient given by $V$, that is near to tribimaximal mixing.
The NY decay $\mathcal{E}^2\to h+\mathcal{E}^1$ is controlled by the
Yukawa coupling, without flavor change and extra suppressions (and
similar for $\tilde h$ and $\tilde P$, although we expect them to be
heavier than the SM Higgs, and for this reason we will neglect them in
the rest of our analysis). This channel is open if $M_e>m_h$,
otherwise it proceeds through a virtual Higgs, giving a three body
decay. The charged Yukawa coupling of the heavy fermion also drives
the CY decay $\mathcal{E}^2\to \tilde h^-+\mathcal{N}^1$, with the
flavor of the final neutrino controlled by $V$. As previously
discussed, at leading order the ratios of the widths of
$\mathcal{E}^2_e,\mathcal{E}^2_\mu$ and $\mathcal{E}^2_\tau$ are given
by: $m_e^2:m_\mu^2:m_\tau^2$. Meaning that the electron-resonance,
$\mathcal{E}^2_e$, is the most stable state in this group.

\begin{figure} \begin{center} \begin{picture}(300,300)(0,0)
\Text(-70,260)[c]{NC:} \ArrowLine(-50,260)(-10,260)
\Photon(-10,260)(20,230){3}{3} \ArrowLine(-10,260)(20,275)
\Vertex(-10,260){2} \Text(-30,275)[c]{$\mathcal{E}^2$}
\Text(-7,235)[c]{$Z$} \Text(26,285)[c]{$\mathcal{E}^1$}
\Text(48,260)[c]{$\simeq$}
	\Vertex(110,260){2} \ArrowLine(70,260)(110,260)
\ArrowLine(110,260)(140,260) \Photon(140,260)(170,230){3}{3}
\ArrowLine(140,260)(170,275) \Vertex(140,260){2}
\DashLine(110,260)(110,280){2} \Text(95,270)[c]{$\et$}
\Text(143,235)[c]{$Z$} \Text(125,270)[c]{$e$} \Text(170,281)[c]{$e$}
\Text(110,285)[c]{\small{$\langle H\rangle$}} \Text(185,260)[c]{$+$}
	\Vertex(245,260){2} \ArrowLine(205,260)(245,260)
\Photon(245,260)(305,230){3}{5} \ArrowLine(245,260)(275,275)
\Vertex(275,275){2} \ArrowLine(275,275)(305,275)
\DashLine(275,275)(275,295){2} \Text(230,270)[c]{$\et$}
\Text(278,235)[c]{$Z$} \Text(260,277)[c]{$\et$} \Text(305,281)[c]{$e$}
\Text(275,300)[c]{\small{$\langle H\rangle$}}
	\Text(-70,135)[c]{CC:} \ArrowLine(-50,135)(-10,135)
\Photon(-10,135)(20,105){3}{3} \ArrowLine(-10,135)(20,150)
\Vertex(-10,135){2} \Text(-30,150)[c]{$\mathcal{E}^2$}
\Text(-7,110)[c]{$W$} \Text(26,160)[c]{$\mathcal{N}^1$}
\Text(48,135)[c]{$\simeq$}
	\Vertex(110,135){2} \ArrowLine(70,135)(110,135)
\ArrowLine(110,135)(140,135) \Photon(140,135)(170,105){3}{3}
\ArrowLine(140,135)(170,150) \Vertex(140,135){2}
\DashLine(110,135)(110,155){2} \Text(95,145)[c]{$\et$}
\Text(143,110)[c]{$W$} \Text(125,145)[c]{$e$} \Text(170,156)[c]{$\nu$}
\Text(110,160)[c]{\small{$\langle H\rangle$}}
	\Text(-70,35)[c]{NY:} \Vertex(-20,35){2}
\ArrowLine(-50,35)(-20,35) \DashLine(-20,35)(10,20){2}
\ArrowLine(-20,35)(10,50) \Text(-35,50)[c]{$\mathcal{E}^2$}
\Text(13,10)[c]{$h$} \Text(13,58)[c]{$\mathcal{E}^1$}
\Text(20,35)[c]{$\simeq$}
	\Vertex(60,35){2} \ArrowLine(30,35)(60,35)
\DashLine(60,35)(90,20){2} \ArrowLine(60,35)(90,50)
\Text(45,45)[c]{$\et$} \Text(93,10)[c]{$h$} \Text(93,56)[c]{$e$}
	\Text(150,35)[c]{CY:} \Vertex(200,35){2}
\ArrowLine(170,35)(200,35) \DashLine(200,35)(230,20){2}
\ArrowLine(200,35)(230,50) \Text(185,50)[c]{$\mathcal{E}^2$}
\Text(233,10)[c]{$\tilde h^-$} \Text(233,58)[c]{$\mathcal{N}^1$}
\Text(240,35)[c]{$\simeq$}
	\Vertex(280,35){2} \ArrowLine(250,35)(280,35)
\DashLine(280,35)(310,20){2} \ArrowLine(280,35)(310,50)
\Text(265,45)[c]{$\et$} \Text(313,10)[c]{$\tilde h^-$}
\Text(313,56)[c]{$\nu$} \end{picture} \caption{Feynman diagrams for
the decay channels of $\mathcal{E}^2\simeq
\et+\mathcal{O}(m_e/M_{\ell,e})$, expanding in powers of mass
insertions.} \label{fig-decays-et} \end{center} \end{figure}
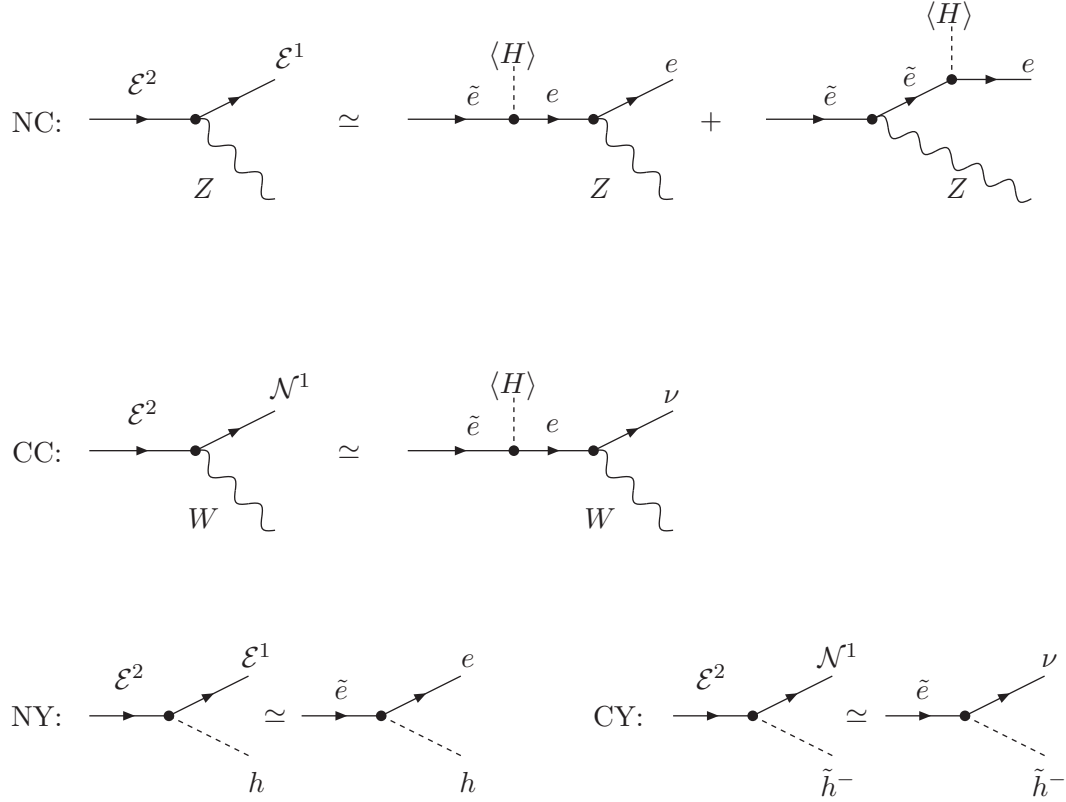

In Fig.~\ref{fig-decays-et} we show the Feynman diagrams corresponding
to the different decay channels of $\mathcal{E}^2_e$ doing a
perturbative expansion in the mass insertions. 

\begin{figure}[h] 
\psfrag{x}[t][b]{{\small $M_{e,\ell}$[GeV]}}
\psfrag{y}[bc][tl]{{\small $\Gamma$[$10^{-11}$GeV]}}
\psfrag{let}{}
\psfrag{et}{}
\psfrag{0.1}{0.1}
\psfrag{1}{1}
\psfrag{10}{10}
\psfrag{200}[c][b]{200}
\psfrag{400}[c][b]{400}
\psfrag{600}[c][b]{600}
\psfrag{800}[c][b]{800}
\psfrag{1000}[c][b]{1000}
\begin{equation}
\begin{array}{cc}
\includegraphics[width=.52\textwidth]{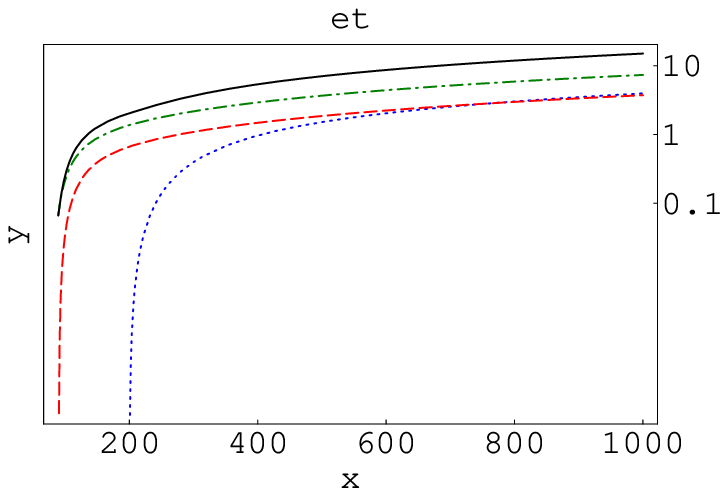}&
\includegraphics[width=.49\textwidth]{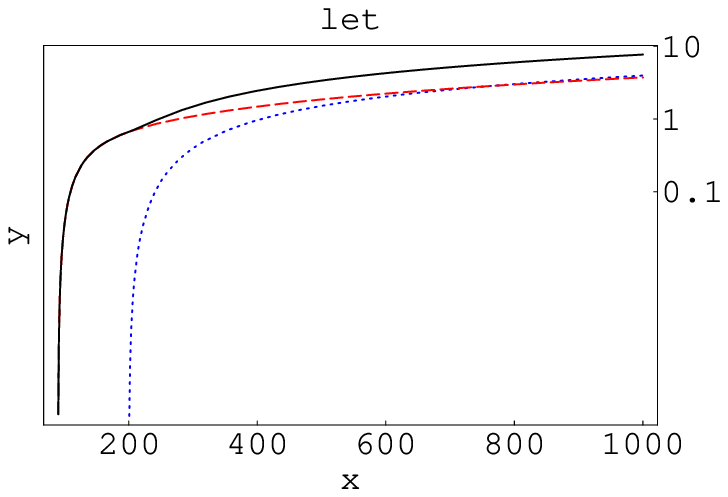} \nonumber\\
(a)&(b) \nonumber
\end{array}
\end{equation}
\caption{(color online) (a) Partial widths of $\mathcal{E}^2_e$ as a
function of $M_e\simeq M_\ell$ and $m_h=200$~GeV.  The
dotted (blue) line corresponds to $\Gamma[\mathcal{E}^2_e\to
h+\mathcal{E}^1_e]$, the dashed (red) line to
$\Gamma[\mathcal{E}^2_e\to Z+\mathcal{E}^1_e]$, the dotted-dashed
(green) line to $\Gamma[\mathcal{E}^2_e\to W+\mathcal{N}^1]$, and the
solid (black) line to the total width. The CC decay dominates the
width.  (b) Partial widths of $\mathcal{E}^3_e$ as a function of
$M_e\simeq M_\ell$ and $m_h=200$~GeV.  The dotted (blue)
line corresponds to $\Gamma[\mathcal{E}^3_e\to h+\mathcal{E}^1_e]$,
the dashed (red) line to $\Gamma[\mathcal{E}^3_e\to
Z+\mathcal{E}^1_e]$ and the solid (black) line to the total width. The
NC decay dominates the width for $M_\ell\lesssim 700$ GeV if $m_h
\gtrsim 200$ GeV.}
\label{plot-width-let}
\end{figure}

In Fig.~\ref{plot-width-let}a we show the partial widths of
$\mathcal{E}^2_e$ as a function of $M_\ell$, for $M_\ell\simeq M_e$,
with $m_h=200$~GeV. We can see that the width is dominated by the
charged current decay and for $M_\ell=300$~GeV (600, 1000 GeV) the
corresponding `lifetime' is
$\tau_{\mathcal{E}^2_e}\simeq2\cdot10^{-14}$ s ($8\cdot10^{-15},
4\cdot10^{-15}$ s).

\subsection{$\mathcal{E}^3$ total width} 

From Table~\ref{table-interactions} we obtain that the decays of
$\mathcal{E}^3$ are similar to those of $\mathcal{E}^2$, analyzed in
the previous section, except that the charged channels have an extra
suppression factor $m_e/M_{\ell,e}$ in the interactions. For this
reason the decay of $\mathcal{E}^3_e$ is dominated by the neutral
decay channels $\mathcal{E}^3_e\to Z+\mathcal{E}^1_e$ and
$\mathcal{E}^3_e\to h+\mathcal{E}^1_e$, and its total with is smaller
than the width of $\mathcal{E}^2_e$. For $m_h=200$ GeV and
$M_\ell\lesssim 700$ GeV the NC channel dominates.  In
Fig.~\ref{plot-width-let}b we show the partial widths of
$\mathcal{E}^3_e$ as a function of $M_\ell$, for $M_\ell\simeq M_e$,
with $m_h=200$~GeV. For $M_\ell=300$ GeV (600, 1000 GeV) the
corresponding `lifetime' is $\tau_{\mathcal{E}^3_e}=4\cdot10^{-14}$ s
($2\cdot10^{-14}, 9\cdot10^{-15}$ s).

There is an important three-body decay channel that could increase the
width of $\mathcal{E}^3$: $\mathcal{E}^3_e\to
W^-\,W^+\,\mathcal{E}^1_\tau$. At a perturbative level, this process
is given by the decay $\lte_e\to\ltnu_a\,W^-\to W^-\, W^+\, \tau$,
with a virtual $\ltnu_a$ and a suppression factor $m_\tau/M_\ell$
(much larger than the usual $m_e/M_\ell$). However, in the MFV
scenario, this process is forbidden by a GIM-like mechanism.

\subsection{$\mathcal{N}^{3,4}$ total width} The decay of
$\mathcal{N}^{3,4}$ through neutral channels is suppressed by a
Majorana mass $\sim m_\nu^2/m_R^2$, as shown in
Table~\ref{table-interactions}. If the small neutrino masses are
generated by the see-saw mechanism, these decay channels have a huge
suppression and can be neglected. On the other hand, the decay through
charged current interactions, $\mathcal{N}^{3,4}_b\to
W+\mathcal{E}^1_\alpha$, with $\alpha=e,\mu,\tau$, is suppressed by
$V^{\dagger \alpha b} m_e^\alpha/M_\ell$. Let us consider the decay of
the heavy neutrinos $\mathcal{N}^{3,4}_1\to W+\mathcal{E}^1_\tau$,
since neither of the matrix elements $V^{\dagger\alpha 1}$ is small,
this channel is suppressed by $m_\tau^2/M_\ell^2$ only, giving rise to
a rather large partial width $\sim 2\cdot10^{-4}$ GeV for $M_\ell\sim
300$ GeV. A similar thing happens with the other mass eigenstates
$\mathcal{N}^{3,4}_{2,3}$. The decay through charged LW-Higgs
interactions is also flavor changing, and for this reason the dominant
channel is proportional to the $\tau$ Yukawa. Thus, due to the large
mass of the $\tau$ and the large flavor mixings in the leptonic sector
of the SM, the neutral heavy leptons have a small `lifetime' $\sim
10^{-25}$ s, and we do not expect them to produce vertex
displacements.

\section{Experimental perspectives}\label{sec-experiment} 

From the previous Section we conclude that the best candidate to
produce an observable wrong vertex displacement is the LW electron
associated to the SU(2) doublet, $\mathcal{E}^3_e\simeq
\lte_e+\mathcal{O}(m_e/M_{\ell,e})$, since it is expected to have the
smallest width, or largest `lifetime'.  For notational simplicity in
the discussion that follows we will refer to the mass eigenstate
$\mathcal{E}^3_e$ and $\mathcal{E}^1_e$ as $\lte_e$ and $e$,
respectively. We study now the production and detection of a pair of
charged LW electrons in colliders like LHC and ILC.  First, we
enumerate a set of conditions we pursue in order to achieve a clear
identification of events arising from the decay of these particles.

\begin{itemize}

\item The LW electrons will be mostly produced in pairs (single
production is suppressed by a factor $m_e/M_{\ell,e}$) via EW
interactions\footnote{We have not taken into account mixings in the
gauge-boson sector at this stage since they turn out to be highly
suppressed $(\mbox{M}^2_{(\mbox{\tiny W,Z})}/\mbox{M}^2_{(\tiny
\tilde{\mbox{W}},\tilde{\mbox{B}})})$ and we do not expect to have significant
variations in the process we are considering.}:
\begin{equation}
\label{production} q\bar
q \rightarrow A,Z,\tilde B,\tilde W^3 \rightarrow \blte_e \lte_e \
.
\end{equation} 

Thus, we require a pair of correlated LW electrons,
meaning that both of them are created in the same single primary
vertex.  Since all the LW interactions are determined by the SM
couplings, the production cross section only depends on the value of
the LW masses.

\item Under the MFV hypothesis, the $\lte_e$ mainly decays --through
neutral interactions-- into an electron and a $Z$ or a Higgs boson,
which in turn can decay into a fermion pair. For a reasonable Higgs
mass $115\, \mbox{GeV}<m_h<300\,\mbox{GeV}$ and a LW mass
$100\,\mbox{GeV}<M_\ell<1000\,\mbox{GeV}$, either the $Z$ channel
dominates the decay or is halved by the Higgs channel.  Assuming a $Z$
channel for the $\lte_e$ decay, Eq.~(\ref{production}) leads to:
\begin{equation}
\label{let-decay} 
\blte_e \lte_e \rightarrow Z e^+
Z e^- \ .  
\end{equation} 
Each $Z$ can decay hadronically or
leptonically, leading to a final state with an electron-positron pair
and in addition: (i) four jets, (ii) two jets and a lepton-antilepton
pair, or (iii) two lepton-antilepton pairs, depending on whether both,
one or none of the $Z$ decays hadronically.

\item We will require that the traces corresponding to one of the
electrons and two of the jets (or one electron and one
lepton-antilepton pair) converge in a vertex well separated from
another vertex defined by the extrapolation of the traces of the
remaining electron and a pair of jets. The relative position of these
secondary vertices in the transverse plane is such that it brings
forward the presence of two wrong vertex displacements and it is
essential in the positive identification of acausal resonances. We
will return to this point below.

\item A fourth requirement is related to the measurement of the
invariant mass corresponding to the decay products of the two LW
electrons.  We will see below that the reconstruction of the
LW-electron mass is a necessary condition to distinguish acausal from
causal resonances.  This can be obtained by measuring the invariant
mass of the three particles emerging from each displaced vertex, as
explained in the previous paragraph.

\end{itemize}

The resulting physical situation corresponds to the one illustrated in
Fig.~\ref{fig:4}. Projecting that picture onto the transverse plane,
we can obtain the unusual position of the vertices in relation to the
traces associated to them: traces of particles that go into the lower
half plane converge from the vertex located in the upper region, and
vice versa (this is what we stated as the essential condition in the
third item). We see that the resulting momentum of the decay products
is directed from the secondary towards the primary vertex.

Returning to the fourth item, we understand why it is necessary to
reconstruct the LW-electron mass. For instance, if it turns out that
some of the product particles is not detected (missing energy), it
could happen that the true resulting momentum of the decay products
points in the opposite direction (i.e., from the primary to the
secondary vertex), as it would happen in the causal case\footnote{This
may occur in decays of supersymmetric particles. For example, if we
consider the process $\tilde{t} \rightarrow t G$, the visible {\it
stop} decay products may recoil against the invisible Goldstino in a
direction towards the primary vertex~\cite{lnip}.}. By measuring the
invariant mass of the decay products it is possible to distinguish
among both situations.  

\subsection{Displaced vertices at the LHC}\label{sec-signal} 
We briefly discuss the production and detection of LW-charged lepton
pairs at the LHC.  As already explained, since we look for displaced
vertices away from the primary vertex, we will require the transverse
displacement to be larger than a reference value that we take as
$\Delta x=20\mu m$. For a particle with mass $M$ and transverse speed
$v_T$, demanding it to travel a distance larger than $\Delta x$
results in the condition $v_T \, \gamma \,\tau >\Delta x$, where
$\gamma$ and $\tau$ are its relativistic factor and lifetime. This is
equivalent to the following condition on the transverse momentum:
$p_T>M \Delta x \Gamma$, where $\Gamma$ is the total width of the
resonance. This rough estimate allows us to obtain an approximate
minimum transverse momentum $p_T$ for the LW resonance as a function
of its mass. In the case of the LW-charged lepton associated to the
SU(2) doublet, $\lte_e$, we obtain the following cuts in its
transverse momentum:
\begin{eqnarray}
\label{cut-let} 
M_\ell=\left\{
\begin{array}{c} 300 \ \rm{GeV}\\ 400 \ \rm{GeV}\\ 500 \ \rm{GeV}
\end{array} \right. 
\qquad \Longrightarrow \qquad 
p_T\gtrsim \left\{
\begin{array}{c} 450 \ \rm{GeV}\\ 980 \ \rm{GeV}\\ 1700 \ \rm{GeV}
\end{array} \right.  
\ \ .  
\end{eqnarray}
\noindent Thus, only for light LW leptons we expect to obtain a
sensible cross section after we demand a minimum distance between the
primary and secondary vertices. 

\begin{figure}[ht] \centering
\psfrag{x}{$p_T(\lte)$ [GeV]}
\psfrag{y}{$\frac{1}{\sigma}\frac{d\,\sigma}{d\,p_T}$ [GeV$^{-1}$]}
\includegraphics[width=.8\textwidth]{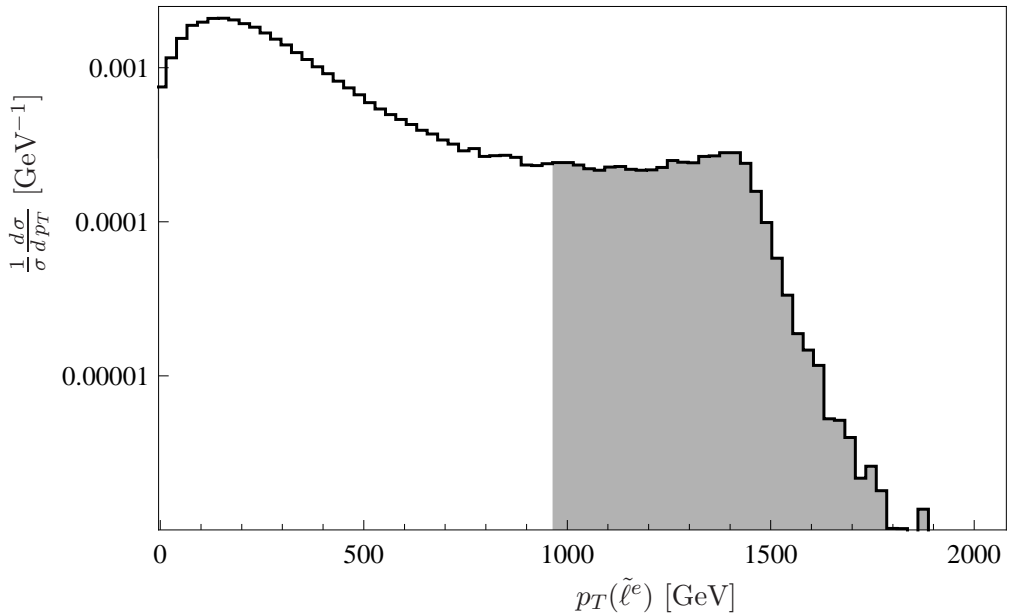}
\caption{$p_T$ distribution for $\blte\lte$ production at the LHC
setting $M_\ell=400$ GeV. The shaded region --which survives after the
cut in $p_T$-- represents the events that would produce wrong vertex
displacements greater than $\Delta x = 20\mu m$. We used $M_{\tilde
W}=M_{\tilde B}=3$ TeV and a center of mass energy of $14$ TeV. The
first resonance corresponds to $Z$ and the second one to $\tilde W^3$
and $\tilde B$, whereas the tail for high $p_T$ is due to the lack
of available energy in the quarks of the proton's beam.}
\label{histograma}
\end{figure}

\begin{figure}[ht] \centering
\psfrag{x}{$M_\ell$ [GeV]}
\psfrag{y}{$\sigma$ [fb]}
\includegraphics[width=.8\textwidth]{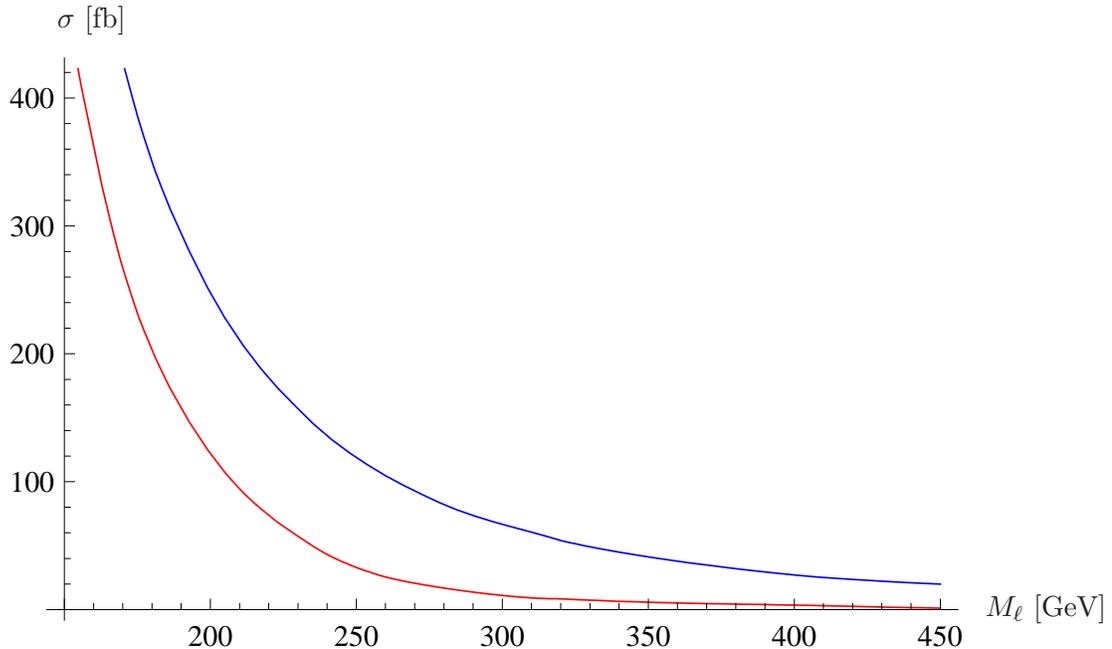}
\caption{Total (upper line) and $p_T$-cut (lower line) cross-sections for
$\blte\lte$ production in LHC. The $p_T$ cut ensures that the wrong vertex
displacement is greater than $\Delta x=20\mu m$. }
\label{x-sections}
\end{figure}

We used MadGraph/MadEvent~\cite{mgme} to obtain the total production
cross section of a LW-charged lepton pair at the LHC with a center of
mass energy of 14 TeV. We have also computed the cross-section after a
cut in $p_T$ such that the transverse vertex displacement is greater
than $\Delta x=20\mu m$. Fig.~\ref{histograma} shows a simulation for
the $p_T$ distribution and its cut in $\blte \lte$ production at the
LHC.  We plot our results for a relevant range of LW masses in
Fig.~\ref{x-sections}, where we have taken $m_h=200\,$GeV.  These
results are slightly suppressed if $m_h$ is lower, see
Table~\ref{table-production-xsect}.  As a general feature we see that
the cross section after the cut has a strong dependence with the LW
scale. For $M_\ell=300$ GeV the total cross section is rather large,
$\sim 66$ fb, and after the cut we still have a sensible cross section
$\sim 11$ fb for $m_h=200\,$GeV and $\sim 8$ fb for $m_h=150\,$GeV. On
the other hand, for $M_\ell\gtrsim 500$ GeV, there is not enough
energy at the LHC to create the highly boosted LW leptons.

\begin{table} \begin{center}
  \begin{tabular}{| c | c | c | c |} \hline $M_\ell$[GeV] & $\sigma_t$[fb]
& $\sigma_{\rm{cut}}|_{m_h=200\,GeV}$[fb]& $\sigma_{\rm{cut}}|_{m_h=150\,GeV}$[fb]  \\ \hline 300 & 66.7 & 11 & 8.4\\ \hline 400 &
27 & 3.5 & 2.6\\ \hline 500 & 15 & -& - \\ \hline \end{tabular}
\end{center} 
\caption{Cross section for the production of a LW-lepton pair at the
LHC, simulated with MadGraph/MadEvent. The second column, $\sigma_t$,
is the total cross section.  The third and fourth columns correspond
to the cross sections after the cut in the transverse momentum for
$m_h=200$ GeV and $m_h=150$ GeV, respectively, needed to obtain an
observable displaced vertex.  The empty box corresponds to a cross
section $\sigma\lesssim{\cal O}(10^{-3})$ fb.}
\label{table-production-xsect} 
\end{table}

\begin{figure}[ht] \centering
\psfrag{x}{$p_T(e)$ [GeV]}
\psfrag{y}{$\frac{1}{\sigma}\frac{d\,\sigma}{d\,p_T}$ [GeV$^{-1}$]}
\includegraphics[width=.8\textwidth]{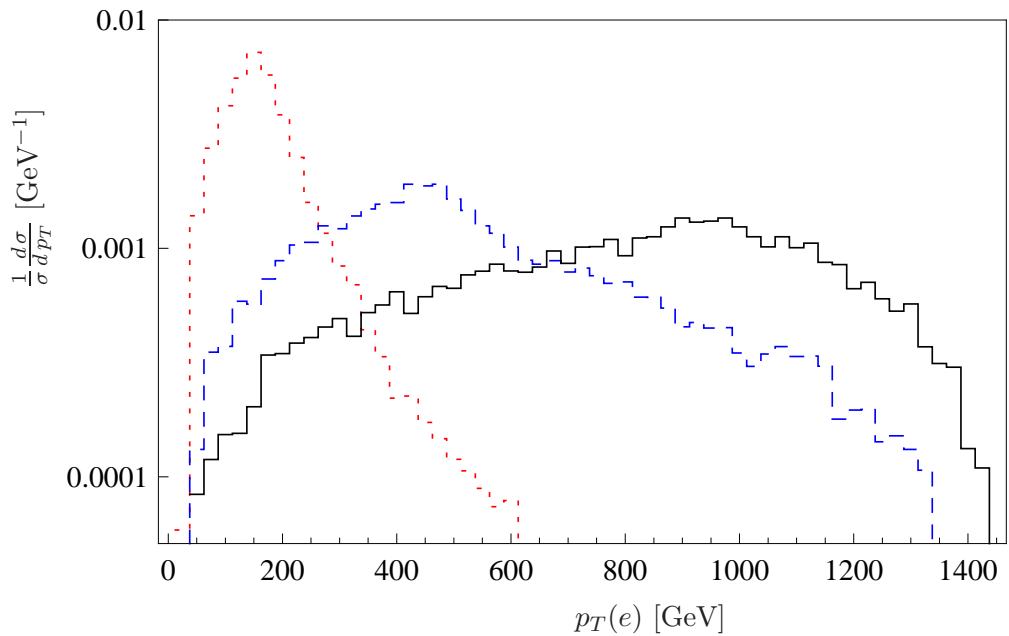}
\caption{(color online) $p_T$ distribution for the most energetic electron
in the signal $e^+e^-jjjj$ after imposing the cut on the $p_T$ of the
LW-leptons to obtain wrong vertex displacements greater than $\Delta x =
20\mu m$. The dotted (red), dashed (blue) and black (solid) lines
correspond to LW masses $M_\ell=200$ GeV$,\ 300$ GeV and $400$ GeV,
respectively.}
\label{pte}
\end{figure}

Once created, each energetic LW lepton will mostly decay into a hard
electron and $Z$ (we will consider $m_h=200$ GeV from now on). The
electrons will have a very large $p_T$ because they are produced in a
two body decay of a heavy LW lepton, and also because the LW state
itself has a large $p_T$.  For instance, for $M_\ell=200$ GeV the
$p_T$ distribution of the final hardest electron is centered in
$p_T\sim200$ GeV, whereas for $M_\ell=400$ GeV is centered in $p_T\sim900$
GeV, see Fig.~\ref{pte}. Therefore, although the cut in the transverse
momentum of the LW lepton suppresses the production cross section, at
the same time, it allows us to impose hard cuts in the $p_T$ of the
most energetic lepton, $p_T\gtrsim \mathcal{O}(200)$ GeV, with low
impact in the signal.  Each $Z$ gauge boson may decay leptonically or
hadronically. The charged leptonic decay leads to a very clean final
state, but has a small branching ratio, giving a cross section that is
too small (unless the LW leptons are very light). Even the case where
only one of the $Z$ bosons decays leptonically may be out of reach,
due to the small production cross section after the
cut~(\ref{cut-let}), see Table~\ref{table-production-xsect}. The
hadronic $Z$ decay dominates, producing a jet pair for each
vector. (Choosing this decay channel reduces the cross section of
Table~\ref{table-production-xsect} by approximately 50\%.)  Therefore,
the signal corresponding to the dominating channel is defined by a
very energetic electron-positron pair and four jets:
\begin{equation}\label{signal}
e^+e^-jjjj \ .  
\end{equation} 
The intermediate $Z$ will also have a
large $p_T$, and could give rise to collinear jets, with small angular
separation. In this case both jets could be resolved as a single jet.
Each pair of jets with an invariant mass corresponding to the $Z$,
when considered together with the proper lepton $e^\pm$, will have an
invariant mass peaked around the LW-lepton mass.
 
The backgrounds for the signal here presented do not seem to cause
further complications.  However, a detailed SM background analysis
should be performed to verify that the whole luminosity for the
$eejjjj$ signal data --with a suitable cut in the electrons'
transverse momentum of $p_T (e) > 200$ GeV-- can be collected to
perform an off-line analysis (this is known as an un-prescaled
trigger).  Afterwards, the peculiarity of this signal would need a
proper vertexing algorithm to cope with the correct wrong vertex
assignment.

\section{Conclusions}\label{sec-conclusions} 

In this article we have defined a new signature called {\it wrong
vertex displacement} which is a distinctive and model independent
signal for new acausal resonances. Microscopic violation of causality
is a feature of the Lee-Wick Standard Model~\cite{Grinstein:2007mp}, a
well motivated model that can solve the hierarchy problem of the
SM. We have proposed to use this new observable to detect acausal LW
resonances at the LHC.

In order to determine whether the LW particles can produce measurable
vertex displacements, we have made a detailed analysis of flavor in
the leptonic sector of the LWSM. We obtained that assuming MFV the
dominant decay channels of the charged LW leptons $\lte_e$ and $\et_e$
are suppressed by the small charged Yukawa couplings, leading to tiny
widths for the first generation of LW electrons (relaxing MFV in the
leptonic sector leads to much larger widths, thus we do not expect to
obtain measurable vertex displacements in that case). For LW lepton
masses of the same order, $M_{\ell} \sim M_{e}$, the best LW candidate
to produce a wrong vertex displacement at the LHC is the LW electron
associated to the SU(2) doublet, $\lte_e$.

The LW electron singlet $\et_e$ could also produce measurable wrong
displaced vertices, although the production cross section is somewhat
smaller due to the larger width compared with the previous
case. The $\et_e$ decays predominantly through CC
interactions to $W\nu_i$, leading to a final state with large missing
energy and making it impossible to reconstruct the mass of the
intermediate resonances.

We have performed a Monte Carlo simulation and computed the
cross-section to pair produce $\lte_e$'s which could generate wrong
vertex displacement greater than $\Delta x = 20\mu m$. Our result is
that for LW-leptonic masses satisfying $M_\ell\lesssim 450$ GeV it
would be possible to observe wrong vertex displacements in the LHC
era. The most promising final state is defined by $e^+e^-jjjj$, with
highly boosted electrons and the jets coming from the hadronic decay
of a pair of $Z$ gauge bosons. This final state allows the
reconstruction of the LW-leptonic masses.  Although a full simulation
analysis is needed, our study remarkably shows that low luminosities
could confirm this signal: demanding five events would require an
integrated luminosity of 1 (3) fb$^{-1}$ for $M_\ell=300$ (400) GeV.
The observation of this signal would point to the existence of acausal
resonances, whereas its non-observation would not rule out models with
this kind of particles, but could put lower bounds on the LW scale and
give us information about the flavor structure of this particular
model.

\begin{acknowledgments} 
We are grateful to Gustavo Otero y Garz\'on and Ricardo Piegaia for
useful discussions on the experimental aspects of this article and
Marcela Carena for valuable information. A.S.\ would like to thank
Georges Azuelos and Hern\'an Wahlberg for helpful communications. The
work of L.D.\ was supported by CONICET and partially supported by the
State of S\~ao Paulo Research Foundation (FAPESP). The work of
C.S. was supported by CONICET and partially supported by the U.~S.
Department of Energy, Office of Nuclear Physics under contract
No. DE-FG02-93ER40756 with Ohio University.
\end{acknowledgments}

\appendix

\section{Diagonalization of the leptonic mass
matrices}\label{Ap-diagonalization} In this section we diagonalize the
leptonic mass matrices. We consider first the sector of charged
leptons, using Eq.~(\ref{Le-flavor-diagonal}) and MFV we can
disentangle the mixings between generations. Using Eq.~(\ref{defE1}) 
\noindent we can write the quadratic Lagrangian as: \begin{equation}
{\cal L}_e=i \bar E \dslash \eta_e E-\bar E_R {\cal M}_e E_L-\bar E_L
{\cal M}^\dagger_e E_R \ , \end{equation} \noindent where
$\eta_e=\rm{diag}(1,-1,-1)$ and \begin{equation}\label{Me} {\cal M}_e=
\left( \begin{array}{ccc} m_e\;&\;0\;&\;-m_e \\ -m_e\;&\;-M_e\;&\;m_e
\\ 0\;&\;0\;&\;-M_\ell \\ \end{array}\right) \ .  \end{equation}
\noindent The independent simplectic rotations $S_{L,R}$ defined by
Eqs.~(\ref{defE1},\ref{defE2},\ref{rotationE}) diagonalize ${\cal
M}_e$ and satisfy the following relations: \begin{eqnarray}
\label{ecs} \mathcal{M}_{e,\mbox{{\tiny phys}}}= S_R^{e\dagger}
\mathcal{M}_e S^e_L \;,\qquad S^e_R \eta_e
S_R^{e\dagger}=\eta_e\;,\qquad S^e_L \eta_e S_L^{e\dagger}= \eta_e\;,
\end{eqnarray} where $\mathcal{M}_{e,\mbox{{\tiny phys}}}$ is the
physical mass matrix, which is diagonal. Expanding in powers of
$\epsilon_{e,\ell}=\frac{m_e}{M_{e,\ell}}\ll 1$ we obtain the following eigenvalues:
\begin{eqnarray}\label{fmasses}
&&m_e[1+\frac{1}{2}(\epsilon_\ell^2+\epsilon_e^2)+\frac{1}{8}(7\epsilon_\ell^4+7\epsilon_e^4+10\epsilon_\ell^2\epsilon_e^2)]+
\mathcal{O}(\epsilon_{\ell,e}^6)\; ,\\
&&M_e[1-\frac{\epsilon_e^2}{2}\frac{M_\ell^2}{M_\ell^2-M_e^2}-
\frac{\epsilon_e^4}{8}\frac{5M_\ell^6-9M_\ell^4M_e^2}{(M_\ell^2-M_e^2)^3}]+
\mathcal{O}(\epsilon_{\ell,e}^6)\; ,\\
&&M_\ell[1+\frac{\epsilon_\ell^2}{2}\frac{M_e^2}{M_\ell^2-M_e^2}+
\frac{\epsilon_\ell^4}{8}\frac{5M_e^6-9M_e^4M_\ell^2}{(M_\ell^2-M_e^2)^3}]+
\mathcal{O}(\epsilon_{\ell,e}^6)\; .
\end{eqnarray}
and for $S^e_{L,R}$ we get:
\begin{eqnarray}\label{Se} S^e_L-1= \left[ \begin{array}{ccc}
\frac{\epsilon_e^2}{2} & -\epsilon_e &
\frac{\epsilon_\ell^4}{\epsilon_e^2-\epsilon_\ell^2} \\ -\epsilon_e &
\frac{\epsilon_e^4(\epsilon_e^2-2\epsilon_\ell^2)}{2(\epsilon_\ell^2-\epsilon_e^2)^2}
& \frac{\epsilon_\ell^2 \epsilon_e}{\epsilon_\ell^2-\epsilon_e^2} \\
-\epsilon_\ell^2 & \frac{\epsilon_\ell^2
\epsilon_e}{\epsilon_e^2-\epsilon_\ell^2} & -\frac{\epsilon_\ell^4
\epsilon_e^2}{2(\epsilon_\ell^2-\epsilon_e^2)^2} \end{array} \right] \ ,
\ \ S^e_R-1= \left[ \begin{array}{ccc} \frac{\epsilon_\ell^2}{2} &
\frac{\epsilon_e^4}{\epsilon_\ell^2-\epsilon_e^2} & -\epsilon_\ell \\
-\epsilon_e^2 &
\frac{\epsilon_\ell^2\epsilon_e^4}{2(\epsilon_\ell^2-\epsilon_e^2)^2} &
\frac{\epsilon_\ell \epsilon_e^2}{\epsilon_\ell^2-\epsilon_e^2} \\
-\epsilon_\ell & \frac{\epsilon_\ell\epsilon_e^2}{\epsilon_e^2-\epsilon_\ell^2}
&
\frac{\epsilon_\ell^4(\epsilon_\ell^2-2\epsilon_e^2)}{2(\epsilon_\ell^2-\epsilon_e^2)^2}
\end{array} \right] \ .  \end{eqnarray} \noindent The solution for the
second and third generations is obtained exchanging the index $e$ by
$\mu$ or $\tau$. This solution is valid for $M_\ell\neq M_e$, the solution for $M_\ell=M_e$ 
can be obtained in a similar way.


We consider now the sector of neutral leptons. Using
Eqs.~(\ref{Lnu-flavor}) and (\ref{rotation-nu}) and imposing MFV we
can disentangle the generation mixing.
From Eq.~(\ref{defN1}) we can write the quadratic Lagrangian as:
\begin{equation} {\cal L}_\nu=\frac{i}{2}\bar N \dslash \eta_\nu
N-\frac{1}{2}\bar N {\cal M}_\nu N \ , \end{equation} \noindent where
$\eta_\nu=\rm{diag}(1,1,-1,-1)$ and \begin{equation}\label{Mnu}
{\mathcal M}_\nu= \left( \begin{array}{cccc} 0\;&\;m_\nu&\;0\;&\;0 \\
m_\nu\;&\;m_R\;&\;-m_\nu&\;0 \\ 0\;&\;-m_\nu&\;0\;&\;-M_\ell \\
0&\;0&\;-M_\ell&\;0 \end{array}\right) \ .  \end{equation} \noindent 
The simplectic rotation matrix $S^\nu$, that diagonalizes ${\mathcal
M}_\nu$, see Eq.~(\ref{rotationN}), satisfies: \begin{eqnarray}
\label{ecsnu} \mathcal{M}_{\nu,\mbox{{\tiny phys}}}= S^{\nu\dagger}
\mathcal{M}_\nu S^\nu \;,\qquad S^\nu \eta_\nu
S^{\nu\dagger}=\eta_\nu\;, \end{eqnarray} where
$\mathcal{M}_{\nu,\mbox{{\tiny phys}}}$ is the physical mass matrix,
which is diagonal. Expanding in powers of $m_\nu$, for $m_\nu\ll
m_R,M_\ell$, we obtain the following eigenvalues: \begin{eqnarray}
\label{ecmnu}
m_{\nu,1}&=&\frac{m_\nu^2}{m_R}-\frac{m_\nu^4}{m_R^3}+\mathcal{O}(m_\nu^6)\
, \\
m_{\nu,2}&=&m_R+m_\nu^2\frac{M_\ell^2}{m_R^2(M_\ell^2-m_R^2)}+m_\nu^4\frac{M_\ell^4(M_\ell^2-3m_R^2)}{m_R^3(m_R^2-M_\ell^2)^3}+\mathcal{O}(m_\nu^6)\
, \\
m_{\nu,3}&=&M_\ell+\frac{m_\nu^2}{2(m_R-M_\ell)}-m_\nu^4\frac{3m_R-5M_\ell}{M_\ell(m_R-M_\ell)^3}+\mathcal{O}(m_\nu^6)\
, \\
m_{\nu,4}&=&M_\ell-\frac{m_\nu^2}{2(m_R+M_\ell)}-m_\nu^4\frac{3m_R+5M_\ell}{M_\ell(m_R+M_\ell)^3}+\mathcal{O}(m_\nu^6)\
, \end{eqnarray} \noindent and: \begin{eqnarray}\label{Snu} S^\nu=
\left[ \begin{array}{cccc} 1-\frac{m_\nu^2}{2m_R^2} &\;
\frac{m_\nu}{m_R} &\; \frac{m_\nu^2}{\sqrt{2}M_\ell(M_\ell-m_R)} &\;
\frac{m_\nu^2}{\sqrt{2}M_\ell(M_\ell+m_R)} \\ -\frac{m_\nu}{m_R} &\;
1-\frac{m_\nu^2M_\ell^2(M_\ell^2-3m_R^2)}{2m_R^2(m_R^2-M_\ell^2)^2}
&\; \frac{m_\nu}{\sqrt{2}(M_\ell-m_R)} &\;
-\frac{m_\nu}{\sqrt{2}(M_\ell+m_R)} \\ -\frac{m_\nu^4}{m_R^2M_\ell^2}
&\; \frac{m_\nu m_R}{m_R^2-M_\ell^2} &\;
-\frac{1}{\sqrt{2}}-\frac{m_\nu^2m_R}{4\sqrt{2}M_\ell(M_\ell-m_R)^2}
&\;
-\frac{1}{\sqrt{2}}+\frac{m_\nu^2m_R}{4\sqrt{2}M_\ell(M_\ell+m_R)^2}
\\ \frac{m_\nu^2}{m_RM_\ell} &\; \frac{m_\nu M_\ell}{m_R^2-M_\ell^2}
&\;
-\frac{1}{\sqrt{2}}+\frac{m_\nu^2(m_R-2M_\ell)}{4\sqrt{2}M_\ell(M_\ell-m_R)^2}
&\;
\frac{1}{\sqrt{2}}+\frac{m_\nu^2(m_R+2M_\ell)}{4\sqrt{2}M_\ell(M_\ell+m_R)^2}
\end{array} \right] \ , \end{eqnarray} \noindent where we have written
only the first non-trivial corrections for every entry of $S^\nu$.
There is a similar solution for every generation, that can be obtained
by considering the corresponding Dirac neutrino mass $m_\nu$.

\section{Interactions between mass eigenstates}\label{Ap-interactions}
In this section we show the leptonic interactions between the mass
eigenstates. From Eqs.~(\ref{LNC},\ref{rotationE},\ref{rotationN}) we
obtain the following neutral current interactions:
\begin{eqnarray}\label{LNC1} \mathcal{L}_{NC}=-(Z_\mu+\tilde Z_\mu)
&[&g_z^{e_L}\bar{\mathcal{E}}_L\gamma^\mu \eta_e \mathcal{E}_L
+g_z^{e_R}\bar{\mathcal{E}}_R\gamma^\mu \eta_e \mathcal{E}_R
\nonumber \\
&+&(g_z^{e_L}-g_z^{e_R})\bar{\mathcal{E}}^i\gamma^\mu(S^{e\dagger}_{L,i2}S^e_{L,2j}P_L-S^{e\dagger}_{R,i3}S^e_{R,3j}P_R)\mathcal{E}^j
\nonumber \\ &+&g_z^{\nu_L}\bar{\mathcal{N}}^i\gamma^\mu
(S^{\nu\dagger}_{i1}S^{\nu}_{1j}P_L-S^{\nu\dagger}_{i3}S^{\nu}_{3j}P_L-S^{\nu\dagger}_{i4}S^{\nu}_{4j}P_R)\mathcal{N}^j]
\ , \end{eqnarray} \noindent where $P_{L,R}$ are the Left and Right
projectors, we have to sum over $i,j=1,2,3$ for the charged leptons
and $i,j=1,\dots 4$ for the neutral leptons, and a sum over a
generation index is understood. We can see explicitly that the
interactions between the charged leptons are not diagonal because
$g_z^{e_L}\neq g_z^{e_R}$. The elements of the matrices $S^{e,\nu}$
can be read form Eqs.~(\ref{Se}) and~(\ref{Snu}).

The neutral Higgs interactions are given by:
\begin{eqnarray}\label{LNY1}
\mathcal{L}_{NY}=&-&\frac{y_e}{\sqrt{2}}\bar{\mathcal{E}}^i_R(S^{e\dagger}_{R,i1}-S^{e\dagger}_{R,i2})(h-\tilde
h+i\tilde P)(S^{e}_{L,1j}-S^{e}_{L,3j})\mathcal{E}^j_L +
\rm{h.c.}\nonumber \\ &-&\frac{y_\nu}{\sqrt{2}}\bar{\mathcal{N}}^i_R
S^{\nu\dagger}_{i2}(h-\tilde h-i\tilde
P)(S^{\nu}_{1j}-S^{\nu}_{3j})\mathcal{N}^j_L+ \rm{h.c.}\ ,
\end{eqnarray} \noindent where a sum over a generation index is
understood.

For the charged current and the charged LW-Higgs interactions we
obtain: \begin{eqnarray}\label{LCC1}
\mathcal{L}_{CC}=&-&\frac{g_2}{\sqrt{2}}(W^+_\mu+\tilde
W^+_\mu)\bar{\mathcal{N}}^i\gamma^\mu V
(S^{\nu\dagger}_{i1}S^e_{L,1j}P_L-S^{\nu\dagger}_{i3}S^{e}_{L,3j}P_L-S^{\nu\dagger}_{i4}S^e_{R,3j}P_R)\mathcal{E}^j
+ \rm{h.c.}\ , \\ \mathcal{L}_{CY}=&
&y_\nu\bar{\mathcal{N}}^i_RS^{\nu\dagger}_{i2}V\tilde
h^+(S^{e}_{L,1j}-S^{e}_{L,3j})\mathcal{E}^j_L + \rm{h.c.}\nonumber \\
&+&y_e\bar{\mathcal{E}}^i_R
(S^{e\dagger}_{R,i1}-S^{e\dagger}_{R,i2})V\tilde
h^-(S^{\nu}_{1j}-S^{\nu}_{3j})\mathcal{N}^j_L+ \rm{h.c.}\ ,
\label{LCY1} \end{eqnarray} \noindent where a sum over generations is
understood.

{}

\end{document}